\documentclass{article}
\usepackage{spconf,amsmath,graphicx,hyperref}
\usepackage{cite}
\usepackage{amsmath,amssymb,amsfonts}
\usepackage{algorithmic}
\usepackage{graphicx}
\usepackage{textcomp}
\usepackage{xcolor}
\usepackage{booktabs}
\usepackage{multirow}
\usepackage{subfloat}
\usepackage{subcaption}

\title{Ensuring Reliable Participation in Subjective Video Quality Tests Across Platforms }
\name{Babak Naderi, Ross Cutler}
\address{Microsoft Corporation, Redmond, USA}
\begin{document}
%
\maketitle
\begin{abstract}

Subjective video quality assessment (VQA) is the gold standard for measuring end-user experience across communication, streaming, and UGC pipelines. Beyond high-validity lab studies, crowdsourcing offers accurate, reliable, faster, and cheaper evaluation-but suffers from unreliable submissions by workers who ignore instructions or game rewards. Recent tests reveal sophisticated exploits of video metadata and rising use of remote-desktop (RD) connections, both of which bias results. We propose objective and subjective detectors for RD users and compare two mainstream crowdsourcing platforms on their susceptibility and mitigation under realistic test conditions and task designs.
\end{abstract}
\section{Introduction}
\label{sec:intro}

Crowdsourcing enables rapid, scalable annotation, easing data bottlenecks for large models. In multimedia subjective quality assessment, progress began with ITU-T P.808 for speech. For video, ITU-T P.910~\cite{itu-t_recommendation_p910_subjective_2023} defines laboratory procedures, and an open-source crowdsourcing adaptation was validated in~\cite{naderi_crowdsourcing_2024-1} with mechanisms to ensure scores' validity and reliability. The P.910 standard includes methods such as Absolute Category Rating (ACR), Degradation Category Rating (DCR), and Comparison Category Rating (CCR).


In ACR, participants view a single video and rate quality on a five-point scale from \textit{Excellent (5)} to \textit{Bad (1)}; scores are aggregated per clip/condition as Mean Opinion Scores (MOS). DCR and CCR are double-stimulus methods showing reference and processed videos. In CCR, participants see both in randomized order and rate the second relative to the first on a seven-point scale, ranging from \textit{Much worse (-3)} through \textit{About the same (0)} and to \textit{Much better (+3)}. Ratings are normalized to reflect processed vs.~reference quality and aggregated per clip/condition as Comparison MOS (CMOS). CCR typically offers higher sensitivity than ACR~\cite{itu-t_recommendation_p910_subjective_2023}.


Low-effort or dishonest responses in crowdsourcing threaten validity. A layered defense is recommended in literature: (i) ex-ante controls (qualifications, screening), (ii) in-task validation (gold questions, attention checks, repeats, time/focus gates), (iii) behavioral analytics (interaction traces, timing), and (iv) ex-post filtering/aggregation (outlier removal, reliability modeling)~\cite{daniel_quality_2018,hirth_analyzing_2013,rzeszotarski_crowdscape_2012}.


For subjective VQA, best practices include environment/device checks (adequate resolution, brightness), qualification tests (color vision, acuity), embedded references and consistency checks (gold stimuli with known answers, trapping stimuli), and monitoring playback completion and viewing time to deter “rate-without-consuming” behavior~\cite{buchholz_crowdsourcing_2011,ribeiro_crowdmos_2011,naderi_effect_2015,hossfeld_best_2014}. With strict screening and post-hoc cleansing, these yield reliable scores consistent with lab studies~\cite{ribeiro_crowdmos_2011,naderi_crowdsourcing_2024-1}.


A major threat is location misrepresentation: workers use Virtual Private Networks (VPNs), Virtual Private Servers (VPSs), or Remote Desktops (RDs) to bypass geo-eligibility. Empirical studies find this common, making IP-based blocking insufficient. Countermeasures combine network signals (ASN/VPN lists, IP–GPS mismatches), cultural/linguistic checks, latency and human reaction-time measures, duplicate-GPS heuristics, and survey-platform server-side blocks \cite{dennis_online_2020,malfait_addressing_2025,kennedy_shape_2020}. To the best of our knowledge, RD detection primarily uses reaction-time testing \cite{malfait_addressing_2025}, which is confounded by network, user responsiveness, and input-device effects.


\section{Methods}
\label{sec:method}


This paper focuses on VQA using Comparison Category Rating (CCR) in crowdsourcing and two suspicious-pattern scenarios: (i) limits of “gold” stimuli—some workers pass controls yet rate inconsistently; (ii) effects of Remote Desktop (RD) use. Our first observations on Amazon Mechanical Turk (AMT), 14\% of raters appeared to connect from non-target countries, and 10\% showed multiple IPs per worker, indicating Virtual Private Network (VPN) or RD usage. VPNs mainly affect demographics, but RD is worse for video quality: RD protocols decode and re-encode videos, adding/removing artifacts that corrupt subjective results.

\subsection{Dataset}
A synthetic dataset was created using 10 talking-head videos from the VCD open-source dataset~\cite{naderi_vcd_2024}, representing user-generated content recorded with diverse devices, environments, participants, and lighting conditions. All source videos have very good quality (MOS = 4.65). The source videos were processed with the degradations listed in Table~\ref{table:deg}, each applied at multiple intensity levels, to generate Processed Video Sequences (PVS).

To analyze suspicious rating patterns, a diagnostic dataset was also constructed with three variations of gold stimuli in addition to the original set (v0). In v1, source and PVS clips were swapped to invert rating polarity. In v2, clips were identical to v0 but renamed with random identifiers. In v3, PVSs were re-encoded with a low Constant Rate Factor (CRF) in H.264 to ensure higher file size and bitrate than the source videos, without changing the perceptual quality.

\begin{table}[t]
\caption{Distortions used in Synthetic Dataset applied on source videos. Number of conditions shows number of intensity levels used.}
\label{table:deg} 
\centering
\resizebox{0.9\columnwidth}{!}{
    \begin{tabular}{l c}
    \toprule
    \textbf{Distortions} & \textbf{N conditions}\\
    \midrule        
Blurring & 7 \\
Scaling (down and up) - Bilinear & 5 \\
JPEG Compression & 10 \\
H.264 Quantization & 6 \\
Frame freezing & 9 \\
JPEG Compression x Scaling & 5x5 \\
Random Combinations + noise and color distortions & 10 \\
\midrule        
Overall & 72 \\

 \bottomrule    
\end{tabular}
}
\end{table}

\subsection{Gold stimuli}
We invited 26 workers who had exhibited suspicious behavior in previous studies to participate in a new test using the diagnostic set. Three participants failed at least one sample in all four variations, while fifteen passed versions v0–v2 but failed in v3, where file sizes were manipulated. This pattern indicates systematic behavior among this subset of workers. Accordingly, we recommend designing gold-standard stimuli such that metadata from the file does not correspond to the expected answer.

A follow-up study was conducted in which two sets of new gold stimuli were included in each session: one used for data cleaning and one as a hidden metric to measure rating accuracy after data cleaning. Additionally, 40 clips were processed with H.264 at three CRF levels (0, 17, 28). In a new subjective test on AMT, the proposed gold questions rejected 38.8\% of submissions. Among the accepted submissions, participants passed the hidden gold questions in 97.5\% of cases. CMOS results for the three CRF levels, obtained with different gold question types, are reported in Table~\ref{table:gold}. Using the proposed gold set, the results indicate that H.264 encoding up to CRF 17 is perceptually lossless, in line with theoretical expectations.

\begin{table}[h]
\caption{Effect of different gold stimuli strategy on CMOS values for three CRF categories.}
\label{table:gold} 
\centering
\resizebox{\columnwidth}{!}{
    \begin{tabular}{l c c c}
    \toprule
    \multirow{2}{*}{\textbf{Gold stimuli}} &   \multicolumn{3}{c}{\textbf{CMOS (95\%CI)}}  \\    
    & \textbf{CRF=0}&  \textbf{CRF=17} &  \textbf{CRF=28}\\
    \midrule        
    No gold stimuli &       0.738 (0.077) & -1.601 (0.126) & -1.618 (0.092) \\
    Original gold set &     -0.016 (0.047) & -1.915 (0.122) & -2.008 (0.080) \\
    Proposed gold set &     0.023 (0.085) & -0.091 (0.124) & -0.188 (0.900) \\
    
 \bottomrule    
\end{tabular}
}
\end{table}

\subsection{RD Participants}
It is generally recommended that applications disable unnecessary graphical effects in RD sessions. Native applications can check registry keys to determine whether they are running in a RD environment and adapt accordingly. Modern web browsers such as Edge and Chrome employ similar mechanisms and communicate this through corresponding CSS settings to webpages. Since the crowdsourcing session are delivered as a webpage, this functionality was used to automatically label sessions as either RD or non–RD. Preliminary tests showed that this approach successfully detected all RD connections to Windows PCs with default settings.

To evaluate the performance of the detection method and the impact of RD usage on perceptual quality in subjective video quality tests, a CCR crowdsourcing experiment was conducted using the synthetic dataset and the P.910-Crowd framework~\cite{naderi_crowdsourcing_2024-1} on the AMT platform. Data cleaning was performed following the procedure in~\cite{naderi_crowdsourcing_2024-1}, resulting in an average of 18.1 valid votes per video clip after the removal of unreliable submissions. In total, 46 workers participated, of whom 67\% were labeled as RD users. Ratings from RD and non–RD participants were aggregated separately by distortion condition, and statistical significance tests were performed to assess whether the two groups rated distortions differently. Table~\ref{table:remote_sig} reports the percentage of cases in which RD users rated quality significantly differently from non–RD users. The results indicate substantial discrepancies between the two groups across a diverse set of degradations.

Figure~\ref{table:remote_sig} illustrates the quality scores in terms of CMOS for two representative cases: blurring and frame freezing. While RD participants generally followed the trend in blurring artifacts, they did not perceive the quality drops as strongly as non–RD participants, likely due to additional encoding/decoding effects introduced by the RD connection. In contrast, RD participants completely failed to detect severe frame-freeze degradations (e.g., 50\% freeze with 5 continues frames). This may be explained either by similar artifacts present in the reference video due to poor RD performance or by participants attributing the distortion to their connection rather than the study itself.

Overall, RD participants provide unreliable scores, as their video quality assessments are confounded by additional processing and network effects introduced by the RD system. Consequently, such participants should be excluded from crowdsourced subjective video quality tests. In a longitudinal observation of 1,834 AMT workers
who were supposedly U.S-based and participated in more than 69,000 video quality assessment sessions in a 6 month period, 68\% of the workers were detected at least once using a RD connection. This demonstrates the dramatic prevalence of RD use in a mainstream crowdsourcing platform.

\begin{table}[t]
\caption{Percentage of cases where RD users rated a distortion significantly different from non-RD participants.}
\label{table:remote_sig} 
\centering
\resizebox{\columnwidth}{!}{
    \begin{tabular}{l c c}
    \toprule
    \textbf{Distortions} & \textbf{N conditions}& \textbf{\% of discrepancy }\\
    \midrule        
Blurring & 7 & 86\%\\
Scaling (down and up) - Bilinear & 5  & 20\%\\
JPEG Compression & 10 & 10\% \\
H.264 Quantization & 6 & 33\% \\
Frame freezing & 9 & 56\% \\
JPEG Compression x Scaling & 5$\times$5 & 44\% \\

 \bottomrule    
\end{tabular}
}
\end{table}

\begin{figure*}[t]
    \centering    
    \subfloat[]{\includegraphics[width=0.45\textwidth]{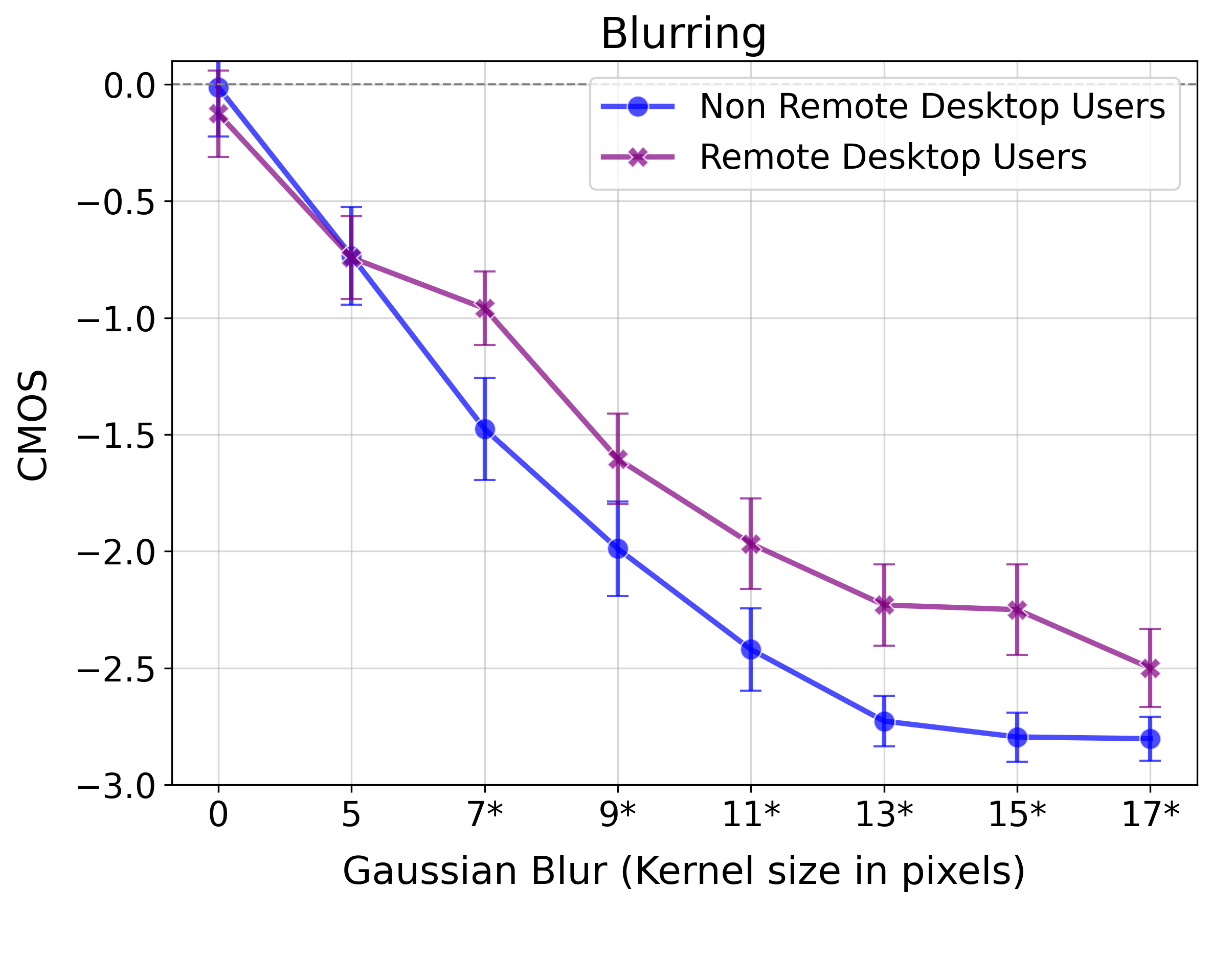}
    \label{fig:blur}}
    \hfill   
    \subfloat[]{\includegraphics[width=0.45\textwidth]{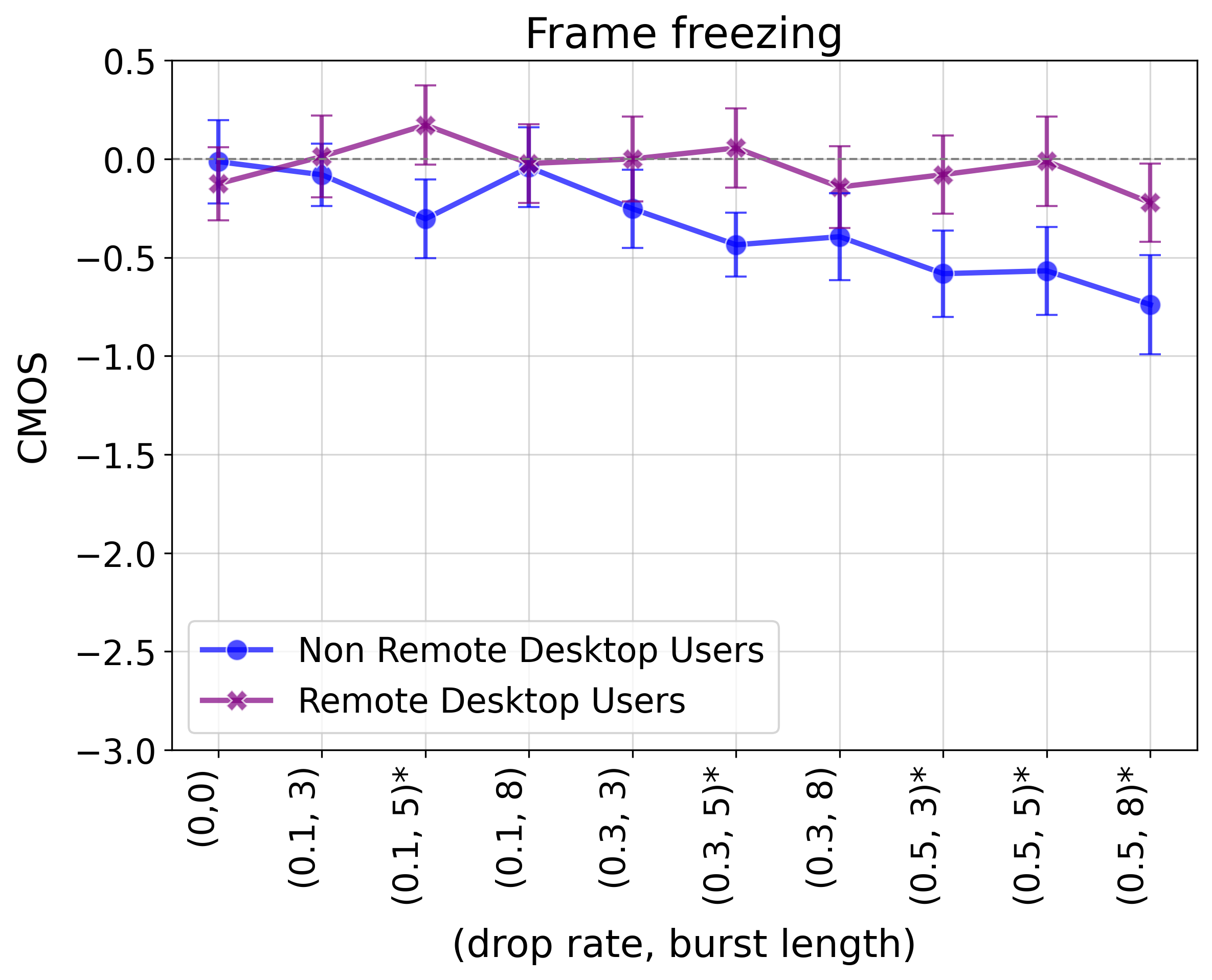}
    \label{fig:frame}}       
    \hfill   
    \caption{Distortion–quality plots for selected degradations in the synthetic dataset aggregated for RD and non-RD participants. Conditions with significant differences between two groups are marked with *.}
    \label{fig:remote_non_remote}    
\end{figure*}

Although the detection method described above is practical, it can be bypassed by changing the default settings of the host machine. Given the substantial impact of RD usage and its prevalence on crowdsourcing platforms, a subjective qualification test was developed to detect RD connections independently of the host settings. In this test, participants are asked to watch two pairs of videos and select which one has better quality.

\subsubsection{Development of Subjective RD Check}
The subjective check must be short to minimize overhead in crowdsourcing video quality sessions. To identify suitable video pairs, multiple experiments were conducted using the CCR method. From the results of the previous experiment, 86 candidate pairs were selected based on two criteria: (1) the clips belonged to the same degradation type, and (2) statistical tests showed contradictory conclusions between RD and non–RD users (i.e., one group judged Clip A significantly better than Clip B, while the other group found no significant difference).

Two additional CCR tests were then carried out to reduce the candidate set, first to 40 pairs and then to 16, using separate groups of RD and non–RD participants. A final CCR test was conducted with 363 participants, of whom 67\% were labeled as RD users using the CSS-based detection method.

Because of the class imbalance, the Synthetic Minority Over-sampling Technique (SMOTE) was applied before training classifiers leading to 47\% and 58\% of data belongs to RD users in training and test set, respectively. Features included the correctness of answers across the 16 video-pair questions and the average percentage of video frames dropped during playback; 20\% of the data were held out for testing. Table~\ref{table:remote_classifier} reports the performance of different classifiers. A Random Forest model was used to identify top-performing features, and a decision tree
was then selected as final model which trained by prioritizing precision.

The top three features included the percentage of dropped frames during playback and two specific video-pair conditions that must be answered correctly. These pairs involved frame-freeze artifacts, at 50\% freeze, 5 consecutive frames, compared against the reference video. The results show that using these indicators, 86\% of participants predicted as non–RD were indeed true non–RD users. 
The code and sample videos for both CCS-based and subjective based RD checks are open-sources in~\cite{naderi_crowdsourcing_2024-1}\footnote{https://github.com/microsoft/P.910}.

\begin{table}[t]
\caption{Performance of different Classifiers with non-RD as positive label.}
\label{table:remote_classifier} 
\centering
\resizebox{\columnwidth}{!}{
    \begin{tabular}{l c c c c c}
    \toprule
    \textbf{Model} &  \textbf{N Features} &\textbf{Accuracy}& \textbf{Precision} & \textbf{Recall} & \textbf{F1-score}\\
    \midrule        
Random Forest  & 17 & 0.90 & 0.89 & 0.95 & 0.92 \\
Support Vector Machine & 17 & 0.79 & 0.80 & 0.84 & 0.82  \\
Decision Tree & 17 & 0.81 & 0.82 & 0.86 & 0.84 \\
\midrule    
Random Forest - reduced & 9 &  0.89 & 0.87 & 0.95 & 0.91 \\
Decision Tree - reduced & 9 & 0.82 & 0.83 & 0.86 & 0.84 \\
\midrule    
Decision Tree (Final) & 3 & 0.74 & 0.86 & 0.65 & 0.74 \\
 \bottomrule    
\end{tabular}
}
\end{table}

\section{Reproducibility}
\label{sec:results}

We applied the above changes on P910-Crowd  framework~\cite{naderi_crowdsourcing_2024-1}, and ran a reproducibility study on AMT and Prolific, repeating each CCR test 5× with independent cohorts. AMT recruited pre-vetted non-RD workers from the longitudinal study; Prolific screened for U.S.-based workers with $\geq$ 98\% approval and $\geq$ 500 submissions, mirroring AMT criteria.

The average number of participants per run, their inter-rater reliabilities, and the correlation coefficients between CMOS values across the N = 5 runs are reported in Table~\ref{table:reprod} for each platform. Both platforms provided similar reproducibility at the condition level. However, Prolific yielded a substantially larger pool of worker groups meeting the requirements of this experiment.
We also observed a significantly lower proportion of RD users on Prolific. In a similar longitudinal study with 3,180 participants, only 3\% were labeled as RD users.

\begin{figure}[h]
    \centering    
    \subfloat[]{\includegraphics[width=0.45\textwidth]{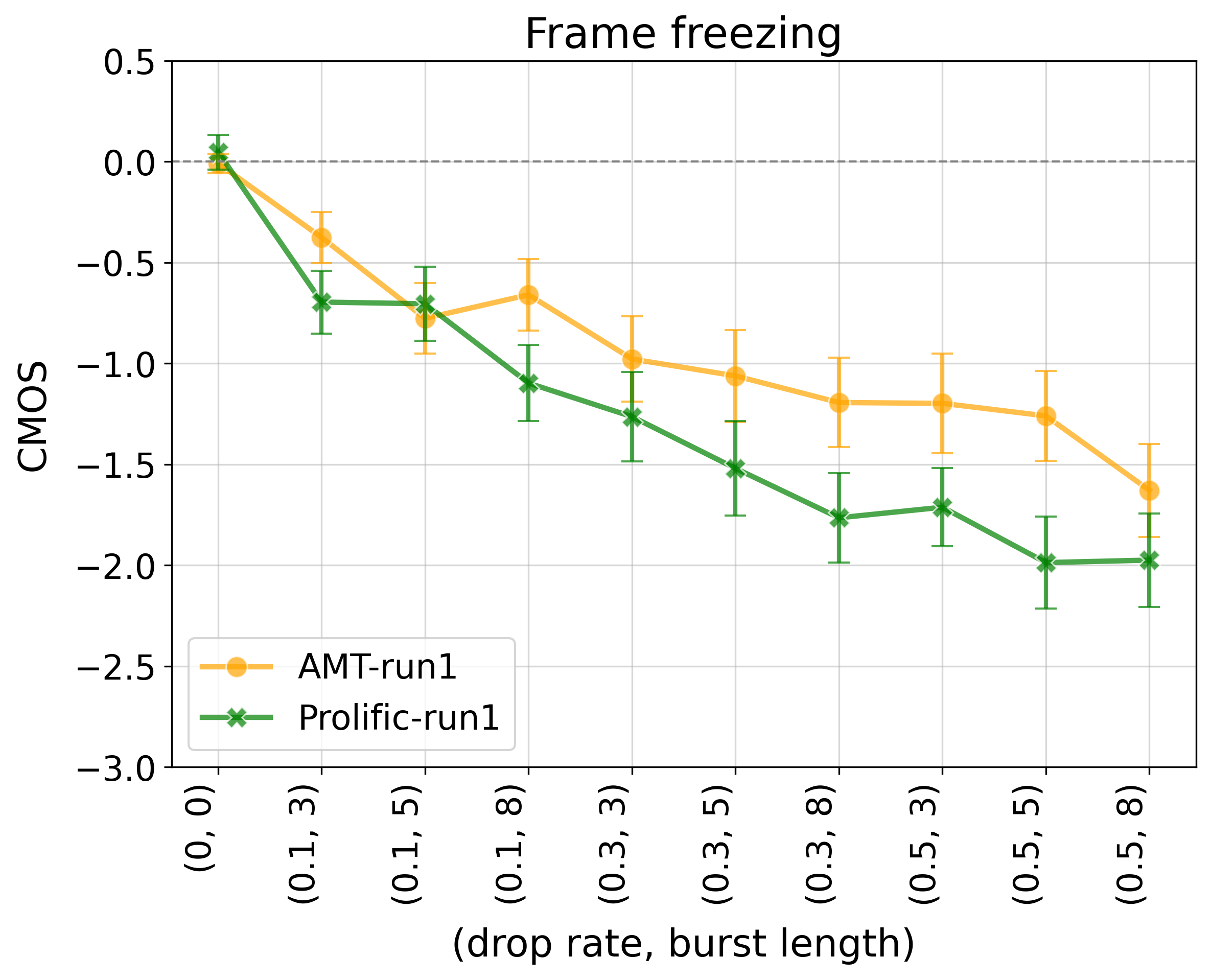}}
    
    \hfill   
    \caption{CMOS values of frame freezing degradations in two runs from the reproducibility tests}
    \label{fig:amt_vs_prolific_frame_freez}    
\end{figure}

Across the two platforms, the average PCC between CMOS values was 0.942 at the clip level and 0.96 at the condition level. For specific degradations, CMOS scores for frame-freezing artifacts were significantly lower on Prolific compared to AMT (see Figure~\ref{fig:amt_vs_prolific_frame_freez}) showing their participants are more sensitive to this distortion.

\begin{table}[t]
\caption{Average correlation coefficients between 5 CCR tests each with separate group of participants on AMT and Prolific platforms.}
\label{table:reprod} 
\centering
\resizebox{\columnwidth}{!}{
    \begin{tabular}{l c c c c }
    \toprule
    & \multicolumn{2}{c}{\textbf{AMT}} & \multicolumn{2}{c}{\textbf{Prolific}} \\
     &  \textbf{Clip} &\textbf{Condition}&  \textbf{Clip} &\textbf{Condition}\\
    \midrule        
    Unique participants & \multicolumn{2}{c}{21} & \multicolumn{2}{c}{157} \\
    Inter Rater Reliability & \multicolumn{2}{c}{0.86} & \multicolumn{2}{c}{0.84} \\
    \midrule
    Pearson  & 0.97 & 0.99 & 0.95 & 0.99 \\
    Spearman & 0.95 & 0.98 & 0.93 & 0.99 \\
    Tau-b & 0.81 & 0.91 & 0.77 & 0.91 \\
    Tau-b 95 & 0.85 & 0.94 & 0.81 & 0.94\\
 \bottomrule    
\end{tabular}
}
\end{table}

\section{Discussion and Conclusion}
\label{sec:discussion}
We reported two current trends that affect VQA test results in crowdsourcing. The first is the ability of malicious participants to pass traditional gold questions by exploiting metadata from video clips, which can be mitigated by the new clip-generation approach introduced in this work. The second is the widespread use of RD connections to access tasks from other regions. We demonstrated the significant impact of RD usage on perceived video quality and participant ratings across a wide variety of degradations typical for video quality tests. We also showed that RD usage is highly prevalent on AMT.
To address this issue, two detection methods —one code-based and one subjective— were introduced. Incorporating these measures enabled high reproducibility of subjective test results across two crowdsourcing platforms. Nevertheless, significant differences in perceived quality for frame-freezing artifacts were observed between platforms which should be investigated in future. Furthermore, while U.S.-based non–RD users represent a limited group on AMT, they are substantially more available on Prolific. This reinforces the necessity of applying the proposed detection checks, as platform populations may change over time.
All proposed methods have been open-sourced to support reliable large-scale subjective data collection within the research community.

\bibliographystyle{IEEEbib}
\bibliography{IC3-AI}

\begin{thebibliography}{10}

\bibitem{itu-t_recommendation_p910_subjective_2023}
ITU-T~Recommendation P.910,
\newblock {\em Subjective video quality assessment methods for multimedia applications},
\newblock International Telecommunication Union, Geneva, Switzerland, 2023.

\bibitem{naderi_crowdsourcing_2024-1}
Babak Naderi and Ross Cutler,
\newblock ``A crowdsourcing approach to video quality assessment,''
\newblock in {\em {ICASSP} 2024-2024 {IEEE} {International} {Conference} on {Acoustics}, {Speech} and {Signal} {Processing} ({ICASSP})}. 2024, pp. 2810--2814, IEEE.

\bibitem{daniel_quality_2018}
Florian Daniel, Pavel Kucherbaev, Cinzia Cappiello, Boualem Benatallah, and Mohammad Allahbakhsh,
\newblock ``Quality control in crowdsourcing: {A} survey of quality attributes, assessment techniques, and assurance actions,''
\newblock {\em ACM Computing Surveys (CSUR)}, vol. 51, no. 1, pp. 1--40, 2018,
\newblock Publisher: ACM New York, NY, USA.

\bibitem{hirth_analyzing_2013}
Matthias Hirth, Tobias Hoßfeld, and Phuoc Tran-Gia,
\newblock ``Analyzing costs and accuracy of validation mechanisms for crowdsourcing platforms,''
\newblock {\em Mathematical and Computer Modelling}, vol. 57, no. 11-12, pp. 2918--2932, 2013,
\newblock Publisher: Elsevier.

\bibitem{rzeszotarski_crowdscape_2012}
Jeffrey Rzeszotarski and Aniket Kittur,
\newblock ``{CrowdScape}: interactively visualizing user behavior and output,''
\newblock in {\em Proceedings of the 25th annual {ACM} symposium on {User} interface software and technology}, 2012, pp. 55--62.

\bibitem{buchholz_crowdsourcing_2011}
Sabine Buchholz and Javier Latorre,
\newblock ``Crowdsourcing {Preference} {Tests}, and {How} to {Detect} {Cheating}.,''
\newblock in {\em Interspeech}, 2011, vol. 2011, p. 12th.

\bibitem{ribeiro_crowdmos_2011}
Flávio Ribeiro, Dinei Florêncio, Cha Zhang, and Michael Seltzer,
\newblock ``{CROWDMOS}: {An} approach for crowdsourcing mean opinion score studies,''
\newblock in {\em 2011 {IEEE} {International} {Conference} on {Acoustics}, {Speech} and {Signal} {Processing} ({ICASSP})}, May 2011, pp. 2416--2419,
\newblock ISSN: 2379-190X.

\bibitem{naderi_effect_2015}
Babak Naderi, Ina Wechsung, and Sebastian \{Möller\},
\newblock ``Effect of being observed on the reliability of responses in crowdsourcing micro-task platforms,''
\newblock in {\em 2015 {Seventh} {International} {Workshop} on {Quality} of {Multimedia} {Experience} ({QoMEX})}. 2015, pp. 1--2, IEEE.

\bibitem{hossfeld_best_2014}
Tobias Hossfeld, Christian Keimel, Matthias Hirth, Bruno Gardlo, Julian Habigt, Klaus Diepold, and Phuoc Tran-Gia,
\newblock ``Best {Practices} for {QoE} {Crowdtesting}: {QoE} {Assessment} {With} {Crowdsourcing},''
\newblock {\em IEEE Transactions on Multimedia}, vol. 16, no. 2, pp. 541--558, Feb. 2014.

\bibitem{dennis_online_2020}
Sean~A Dennis, Brian~M Goodson, and Christopher~A Pearson,
\newblock ``Online worker fraud and evolving threats to the integrity of {MTurk} data: {A} discussion of virtual private servers and the limitations of {IP}-based screening procedures,''
\newblock {\em Behavioral Research in Accounting}, vol. 32, no. 1, pp. 119--134, 2020,
\newblock Publisher: American Accounting Association.

\bibitem{malfait_addressing_2025}
Ludovic Malfait, Neel Chaudhari, and Doh-Suk Kim,
\newblock ``Addressing {VPN} and {VPS} users when conducting subjective tests on crowdsourcing platforms,''
\newblock {\em Quality and User Experience}, vol. 10, no. 1, pp. 1--12, 2025,
\newblock Publisher: Springer.

\bibitem{kennedy_shape_2020}
Ryan Kennedy, Scott Clifford, Tyler Burleigh, Philip~D Waggoner, Ryan Jewell, and Nicholas~JG Winter,
\newblock ``The shape of and solutions to the {MTurk} quality crisis,''
\newblock {\em Political Science Research and Methods}, vol. 8, no. 4, pp. 614--629, 2020,
\newblock Publisher: Cambridge University Press.

\bibitem{naderi_vcd_2024}
Babak Naderi, Ross Cutler, Nabakumar~Singh Khongbantabam, Yasaman Hosseinkashi, Henrik Turbell, Albert Sadovnikov, and Quan Zou,
\newblock ``{VCD}: {A} {Video} {Conferencing} {Dataset} for {Video} {Compression},''
\newblock in {\em {ICASSP} 2024-2024 {IEEE} {International} {Conference} on {Acoustics}, {Speech} and {Signal} {Processing} ({ICASSP})}. 2024, pp. 3970--3974, IEEE.

\end{thebibliography}

\end{document}